\documentclass[12pt]{article}

\usepackage{amsthm}
\theoremstyle{plain}

\theoremstyle{definition}

\theoremstyle{proposition}

\theoremstyle{lemma}

\theoremstyle{remark}

\usepackage{amsmath}
\usepackage{amssymb}
\usepackage[pdftex]{graphicx}

\makeatletter
 
  \@addtoreset{equation}{section}
 \makeatother

\def\Eq#1{\begin{equation} #1 \end{equation}}
\def\Eqr#1{\begin{eqnarray} #1 \end{eqnarray}}
\def\Eqrsubl#1#2{\begin{subequations}\label{#1}\Eqr{#2}\end{subequations}}

\newcommand{\nn}{\nonumber}
\newcommand{\pd}{\partial}

\newcommand{\bea}{\begin{eqnarray}}
\newcommand{\eea}{\end{eqnarray}}

\def\X5sp{{\rm X}_5}
\def\Y3sp{{\rm Y}_3}
\def\Z3sp{{\rm Z}_3}

\begin{document}
\setlength{\oddsidemargin}{0cm}
\setlength{\baselineskip}{7mm}

\begin{titlepage}
\begin{flushright}   \end{flushright} 

~~\\

\vspace*{0cm}
    \begin{Large}
       \begin{center}
         {Path integrals of perturbative strings on curved backgrounds from string geometry theory}
       \end{center}
    \end{Large}
\vspace{1cm}

\begin{center}
           Matsuo S{\sc ato},$^{*}$\footnote
           {
e-mail address : msato@hirosaki-u.ac.jp}
Yuji S{\sc ugimoto}$^{\dagger}$\footnote
           {
e-mail address : yujisugimoto@postech.ac.kr}
and  Kunihito U{\sc zawa}$^{*}$\footnote
           {
e-mail address : kunihito.uzawa@hirosaki-u.ac.jp}\\
      \vspace{1cm}
       
         {$^{*}$\it Graduate School of Science and Technology, Hirosaki University\\ 
 Bunkyo-cho 3, Hirosaki, Aomori 036-8561, Japan}\\

{$^{\dagger}$\it Department of Physics, POSTECH, Pohang 37673, Korea}

\end{center}

\hspace{5cm}

\begin{abstract}
\noindent
String geometry theory is one of the candidates of the non-perturbative formulation of string theory. In this paper, from the closed bosonic sector of string geometry theory,  we derive path integrals of perturbative strings  on all the string backgrounds, $G_{\mu\nu}(x)$, $B_{\mu\nu}(x)$, and $\Phi (x)$, by considering fluctuations around the string background configurations, which are parametrized by the string backgrounds.

\end{abstract}

\vfill
\end{titlepage}
\vfil\eject

\setcounter{footnote}{0}

\section{Introduction}\label{intro}
\setcounter{equation}{0}
String geometry theory is one of the candidates of non-perturbative formulation of string theory. It is formulated by  a   path integral of string manifolds, which belong to a class of infinite-dimensional manifolds, string geometry \cite{Sato:2017qhj}. String manifolds are defined by patching open sets of the model space defined by introducing a topology to a set of strings.  One of the remarkable facts concerning string geometry theory is that the path integral of   perturbative superstrings on the flat background is derived including the moduli of super Riemann surfaces, by considering fluctuations around the flat background in the theory \cite{Sato:2017qhj, Sato:2019cno,  Sato:2020szq}.

Moreover, configurations of fields in string geometry theory include all configurations of fields in the ten-dimensional supergravities, namely string backgrounds \cite{Honda:2020sbl, Honda:2021rcd}. Especially, it is shown that an infinite number of equations of motion of string geometry theory are consistently truncated to finite numbers of equations of motion of the supergravities. That is, string geometry theory includes string backgrounds not as external fields like the perturbative string theories. Dynamics of string backgrounds are a part of  dynamics of  the fields in the theory. It is natural to expect to derive the path integral of perturbative strings on the sting backgrounds by considering fluctuations around the corresponding configurations in string geometry theory. 
Furthermore, a string background that minimizes the energy of the string background configurations, will be chosen spontaneously, because string geometry theory is formulated non-perturbatively \cite{Honda:2020sbl, Honda:2021rcd}.  

For each background, one theory is formulated in case of a perturbative string theory, whereas perturbative string theories not only on the flat background but also on non-trivial backgrounds should be derived from a single theory in case of the non-perturbative formulation of string theory. 
In this paper, from the closed bosonic sector of string geometry theory, we derive the path integrals of  perturbative strings on all the string backgrounds $G_{\mu\nu}(x)$, $B_{\mu\nu}(x)$, and $\Phi (x)$.

The organization of the paper is as follows. In section 2, we briefly review the closed bosonic sector in string geometry theory. In section 3,  we set string background configurations parametrized by the string backgrounds $G_{\mu\nu}(x)$, $B_{\mu\nu}(x)$, and $\Phi (x)$, and  set the classical part of fluctuations representing strings. In section 4, we consider two-point correlation functions of the quantum part of the fluctuations and derive the path integrals of the perturbative strings on the string backgrounds. In section 5, we conclude and discuss our results. In the appendix, we obtain a Green function on the flat string manifold.

\vspace{1cm}

\section{Review of closed bosonic sector in string geometry theory}
\setcounter{equation}{0}
In this paper, we discuss only the closed bosonic sector of string geometry theory. One can generalize the result in this paper to the full string geometry theory in the same way as in \cite{Sato:2017qhj}. The closed bosonic sector \cite{Honda:2020sbl, Honda:2021rcd} is described by a partition function
\begin{align}
Z=\int \mathcal{D}G \mathcal{D}\phi  \mathcal{D}B e^{-S},
\end{align}
where the action is given by
\Eqr{
S&=&
\int {\cal D}\bar{\tau}\, {\cal D}\bar{h}\,{\cal D}X(\bar{\tau})
\sqrt{-G}\,e^{-2\phi}
\left[R+4\nabla_I\phi\nabla^I \phi
-\frac{1}{2}|H|^2 \right],
  \label{action}
}
where $|H|^2=\frac{1}{3!}H_{MNP}H^{MNP}$.
The path integral is defined by integrating a metric $G_{IJ}$, a scalar $\phi$, and a two-form $B_{IJ}$ defined on an infinite dimensional manifold, so-called string manifold. String manifold is constructed by patching open sets in string model space $E$, whose definition is summarized as follows.  First, a global time $\bar{\tau}$ is defined canonically and uniquely on a Riemann surface $\bar{\Sigma}$ by the real part of the integral of an Abelian differential uniquely defined on $\bar{\Sigma}$ \cite{KricheverNovikov1, KricheverNovikov2}.
We restrict $\bar{\Sigma}$ to a $\bar{\tau}$ constant line and obtain $\bar{\Sigma}|_{\bar{\tau}}$. An embedding of $\bar{\Sigma}|_{\bar{\tau}}$ to $\mathbb{R}^{d}$ represents a many-body state of strings in $\mathbb{R}^{d}$, and is parametrized by coordinates $(\bar{h}, X (\bar{\tau}), \bar{\tau})$\footnote{`` $\bar{}$ " represents a representative of the diffeomorphism and Weyl transformations on the worldsheet. Giving a Riemann surface $\bar{\Sigma}$ is equivalent to giving a  metric $\bar{h}$ up to diffeomorphism and Weyl transformations.} where $\bar{h}$ is a metric on  $\bar{\Sigma}$  and $X (\bar{\tau})$ is a map from  $\bar{\Sigma}|_{\bar{\tau}}$ to $\mathbb{R}^{d}$.  String model space $E$  is defined by the collection of the string states by considering all the  $\bar{\Sigma}$, all the values of $\bar{\tau}$, and all the $X (\bar{\tau})$. How near the two string states is defined by how near the values of $\bar{\tau}$ and how near $X (\bar{\tau})$. $\bar{h}$ is a discrete variable in the topology of string geometry, where an $\epsilon$-open neighborhood of $[\bar{h}, X_s(\bar{\tau}_s), \bar{\tau}_s]$ is defined by
\begin{eqnarray}
U([\bar{h}, X_s(\bar{\tau}_s), \bar{\tau}_s], \epsilon)
:=
\left\{[\bar{h},  X(\bar{\tau}), \bar{\tau}] \bigm| \sqrt{|\bar{\tau}-\bar{\tau}_s|^2+  \| X(\bar{\tau})-X_s(\bar{\tau}_s) \|^2} <\epsilon   \right\}.
\label{neighbour}
\end{eqnarray}
As a result, $d \bar{h}$  cannot be a part of basis that span the cotangent space in (\ref{cotangen}), whereas fields are functionals of $\bar{h}$ as in (\ref{LineElement}). The precise definition of the string topology is given in the section 2 in \cite{Sato:2017qhj}. By this definition, arbitrary two string states on a connected  Riemann surface in $E$ are connected continuously. Thus, there is an one-to-one correspondence between a Riemann surface in $\mathbb{R}^{d}$ and a curve  parametrized by $\bar{\tau}$ from $-\infty$ to $\infty$ on $E$. That is, curves that represent asymptotic processes on $E$ reproduce the right moduli space of the Riemann surfaces in $\mathbb{R}^{d}$. Therefore, a string geometry model possesses all-order information of the perturbative string theory.  Indeed, the path integral of perturbative strings on the flat spacetime is derived from the string geometry theory as in \cite{Sato:2017qhj, Sato:2020szq}.
We use the Einstein notation for the index $I$, where $I=\{d,(\mu \bar{\sigma}) \}$.
The cotangent space is spanned by 
\begin{eqnarray}
d X^{d} &:=& d \bar{\tau}, \nonumber \\
d X^{(\mu \bar{\sigma})  }&:=& d X^{\mu} \left( \bar{\sigma}, \bar{\tau} \right), \label{cotangen}
\end{eqnarray}
for $\mu=0,1, \dots, d-1$, while $d \bar{h}_{mn}$ with $m,n=\bar{\tau},\bar{\sigma}$ cannot be a part of the basis because $\bar{h}_{mn}$ is treated as a discrete valuable in the string topology. 
The summation over $\bar{\sigma}$ is defined by $\int d\bar{\sigma}  \bar{e} (\bar{\sigma}, \bar{\tau})$, where $\bar{e}:=\sqrt{\bar{h}_{ \bar{\sigma} \bar{\sigma}}}$. This summation is transformed as a scalar under $\bar{\tau} \mapsto \bar{\tau}'(\bar{\tau},  X(\bar{\tau}))$, and invariant under $\bar{\sigma} \mapsto \bar{\sigma}'(\bar{\sigma})$.

From these definitions, we can write down the general form of the metric of the string geometry as follows.
\begin{align}
ds^2 &(\bar{h}, X(\bar{\tau}), \bar{\tau}) 
\nonumber \\
=
&G_{dd} (\bar{h}, X(\bar{\tau}), \bar{\tau}) (d\bar{\tau})^2
+2 d\bar{\tau} \int d\bar{\sigma}  \bar{e} (\bar{\sigma}, \bar{\tau})  \sum_{\mu}  G_{d \; (\mu \bar{\sigma})}(\bar{h}, X(\bar{\tau}), \bar{\tau}) d X^{\mu}(\bar{\sigma}, \bar{\tau}) \nonumber \\
&+\int d\bar{\sigma}   \bar{e} (\bar{\sigma}, \bar{\tau}) \int d\bar{\sigma}' \bar{e} (\bar{\sigma}', \bar{\tau})  \sum_{\mu, \mu'} G_{ \; (\mu \bar{\sigma})  \; (\mu' \bar{\sigma}')}(\bar{h}, X(\bar{\tau}), \bar{\tau}) d X^{\mu}(\bar{\sigma}, \bar{\tau}) d X^{\mu'}(\bar{\sigma}', \bar{\tau}). \label{LineElement} 
\end{align}
The inverse metric $G^{IJ}(\bar{h}, X_{\hat{D}_{T}}(\bar{\tau}), \bar{\tau})$ is defined by $G_{IJ}G^{JK}=G^{KJ}G_{JI}=\delta^{K}_{I}$, where $\delta^{d}_{d}=1$ and $\delta^{(\mu' \bar{\sigma}')}_{(\mu \bar{\sigma})}=\frac{1}{\bar{e}(\bar{\sigma},\bar{\tau})} \delta^{\mu'}_{\mu} \delta(\bar{\sigma} - \bar{\sigma}')$.
In the following, we use $D:=\int d\bar{\sigma} \bar{e} \delta^{(\mu \bar{\sigma})}_{(\mu \bar{\sigma})}=2\pi d \delta(0)$, then $\delta^M_M=D+1$. Although $D$ is infinity, we treat $D$ as regularization parameter and will take $D \to \infty$ later.
\vspace{1cm}


\section{
String background configurations and fluctuations representing strings}

In this paper, we consider only static configurations, including quantum fluctuations:
\begin{eqnarray}
\partial_d G_{MN}&=&0,  \nonumber \\
\partial_d B_{MN}&=&0,  \nonumber \\
\partial_d \phi&=&0. \label{static}
\end{eqnarray}
In this section, we will set classical backgrounds including string backgrounds and consider fluctuations that represent strings around them.
The Einstein equation of the action (\ref{action}) is given by 
\Eqr{
&&\bar{R}_{MN}
-\frac{1}{4}\bar{H}_{MAB}{\bar{H}_N}^{AB}+2\bar{\nabla}_M\bar{\nabla}_N\bar{\phi}\nn\\
&&~~~~~-\frac{1}{2}{\bar{G}}_{MN}\left(\bar{R}-4\bar{\nabla}_I\bar{\phi}\bar{\nabla}^I\bar{\phi}
+4\bar{\nabla}_I\bar{\nabla}^I\bar{\phi}-\frac{1}{2}\bar{H}^2\right)=0,
    \label{Ein}
}
where $\bar{R}$, $\bar{R}_{MN}$, ${\bar{R}^M}_{NPQ}$, and $\bar{\nabla}_M$ denote the Ricci scalar, Ricci tensor, curvature tensor and covariant derivative constructed from the metric  $\bar{G}_{MN}$, respectively. 
We consider a perturbation with respect to the metric $\bar{G}_{MN}$:
\Eqr{
\bar{G}_{MN}=\hat{G}_{MN}+\bar{h}_{MN},
   \label{Gh}
}
where $\bar{h}_{MN}$ denotes a fluctuation 
around the 0-th order background $\hat{G}_{MN}$.We raise and lower the indices by $\hat{G}_{MN}$ in the following. 
We also consider a perturbation with respect to the 2-form $\bar{B}_{MN}$ and the scalar $\bar{\phi}$ around the 0-th order backgrounds $0$. 

First, we generalize the harmonic gauge to the one when we have the dilaton.   
If we define $\bar{\psi}_{MN}$ as
\Eqr{
\bar{\psi}_{MN}=\bar{h}_{MN}
-\frac{1}{2}\hat{G}^{IJ}
\bar{h}_{IJ}\hat{G}_{MN}
+\Lambda \hat{G}_{MN}\bar{\phi},
  \label{spsi}
}
the Einstein equation (\ref{Ein}) is expressed as 
\Eqr{
&&\hat{R}_{MN}-\frac{1}{2}\hat{G}_{MN}\hat{R}
+\frac{1}{2}\left(-\hat{\nabla}_I\hat{\nabla}^I
\bar{\psi}_{MN}
+\hat{R}_{MA}\,{\bar{\psi}^A}_N
+\hat{R}_{NA}\,{\bar{\psi}_M}^A
-2\hat{R}_{MANB}\,\bar{\psi}^{AB}\right.\nn\\
&&\left.~~~~~
+\hat{\nabla}_M\hat{\nabla}_A
{\bar{\psi}^A}_N
+\hat{\nabla}_N\hat{\nabla}_A
{\bar{\psi}_M}^A
-\hat{\nabla}^I\hat{\nabla}^J
\bar{\psi}_{IJ}\,
\hat{G}_{MN}
+\hat{R}^{IJ}\,
\bar{\psi}_{IJ}\,\hat{G}_{MN}
-\hat{R}\,\bar{\psi}_{MN}\right)\nn\\
&&~~~~~+\left(2-\Lambda\right)
\hat{\nabla}_M\hat{\nabla}_N\bar{\phi}
-\left(2-\Lambda\right)
\hat{G}_{MN}\hat{\nabla}_I\hat{\nabla}^I\bar{\phi}
=0,
   \label{spEin}
}
up to the first order in the fields, $\bar{h}_{IJ}$, $\bar{B}_{IJ}$, and $\bar{\phi}$. 
$\hat{R}$, $\hat{R}_{MN}$, 
${\hat{R}^M}_{NPQ}$, $\hat{\nabla}_M$
denote the Ricci scalar, Ricci tensor, curvature tensor and covariant derivative constructed from the metric $\hat{G}_{MN}$. 
We set $\Lambda=2$ so that the Einstein equation becomes only for $\bar{\psi}_{MN}$.
$\bar{h}_{MN}$ is inversely expressed as 
\Eqr{
\bar{h}_{MN}
&=&\bar{\psi}_{MN}
+\frac{1}{D-1}\left(-\hat{G}^{PQ}
\,\bar{\psi}_{PQ}
+4\bar{\phi}\right)
\hat{G}_{MN}. \label{inverse}
}
We impose a generalization of the harmonic gauge: 
\Eqr{
\hat{\nabla}^M\bar{\psi}_{MN}=0,
    \label{nsh}
}
which reduces to the ordinary harmonic gauge if the dilaton is zero.
Then, the Einstein equation (\ref{spEin}) becomes 
\Eqr{
&&\hat{R}_{MN}-\frac{1}{2}\hat{G}_{MN}\hat{R}
+\frac{1}{2}\left(-\hat{\nabla}_I\hat{\nabla}^I
\bar{\psi}_{MN}
+\hat{R}_{MA}\,{\bar{\psi}^A}_N
+\hat{R}_{NA}\,{\bar{\psi}_M}^A
-2\hat{R}_{MANB}\,\bar{\psi}^{AB}\right.\nn\\
&&\left.~~~~~
+\hat{R}^{IJ}\,
\bar{\psi}_{IJ}\,\hat{G}_{MN}
-\hat{R}\,\bar{\psi}_{MN}\right)
=0.
   \label{nspEin2-2}
}
Next, we set the 0-th order background $\hat{G}_{MN}$ as a flat background:
\Eqr{
\label{metric}
&&\hat{G}_{MN}=a_M\,\eta_{MN}, \label{flatmetric}
}
where $a_d=1$ and $a_{(\mu\bar{\sigma})}=\frac{\bar{e}^3(\bar{\sigma})}
{\sqrt{\bar{h}(\bar{\sigma})}}$. Then, the gauge fixing condition (\ref{nsh}) becomes
\Eq{
\int d\bar{\sigma}\,\bar{e}\,
\pd^{(\mu\bar{\sigma})}
\,\bar{\psi}_{(\mu\bar{\sigma})M}=0, \label{GaugeMetric}
}
the Einstein equation (\ref{nspEin2-2})  becomes Laplace equation,
\Eqr{
\int d\bar{\sigma}\,\bar{e}\,
\pd_{(\mu\bar{\sigma})}\pd^{(\mu\bar{\sigma})}
\,\bar{\psi}_{MN}=0,
   \label{nspEin3}
}
and 
the components of (\ref{inverse}) read
\Eqr{
\bar{h}_{dd}&=&\frac{D-2}{D-1}\,\bar{\psi}_{dd}
+\frac{1}{D-1}\int d\bar{\sigma}''\,
\bar{e}''\,
{\bar{\psi}^{(\mu''\bar{\sigma}'')}}_{(\mu''\bar{\sigma}'')}
-\frac{4}{D-1}\,\bar{\phi},  \nonumber\\
\bar{h}_{d(\mu\bar{\sigma})}&=&\bar{\psi}_{d(\mu\bar{\sigma})},  \nonumber \\
\bar{h}_{(\mu\bar{\sigma})(\mu'\bar{\sigma}')}&=&
\bar{\psi}_{(\mu\bar{\sigma})(\mu'\bar{\sigma}')}
+\frac{\bar{e}^3}
{\sqrt{\bar{h}}}\,
\delta_{(\mu\bar{\sigma})(\mu'\bar{\sigma}')}
\left(\frac{1}{D-1}\,\bar{\psi}_{dd}
\right.\nn\\
&&\left.~~~~~
-\frac{1}{D-1}\int d\bar{\sigma}''\,
\bar{e}''\,
{\bar{\psi}^{(\mu''\bar{\sigma}'')}}_{(\mu''\bar{\sigma}'')}
+\frac{4}{D-1}\bar{\phi}\right).\label{compo}
}

Next, the equation of motion of the scalar of the action \eqref{action} 
\Eqr{
\bar{R}-4\bar{\nabla}_M\bar{\phi}\bar{\nabla}^M\bar{\phi}
+4\bar{\nabla}_M\,\bar{\nabla}^M\bar{\phi}-\frac{1}{2}|\bar{H}|^2=0,
   \label{deq}
}
is written as 
\Eqr{
\hat{R}+\hat{\nabla}^M\hat{\nabla}^N 
\bar{h}_{MN}
-\hat{\nabla}^M\hat{\nabla}_M {\bar{h}^N}_N
+4\hat{G}^{MN}\hat{\nabla}_M
\hat{\nabla}_N\bar{\phi}=0,
   \label{deq1}
}
up to the first order in the fields, $\bar{h}_{IJ}$, $\bar{B}_{IJ}$, and $\bar{\phi}$. 
Furthermore, this can be written as
\Eq{
\int d\bar{\sigma}\,\bar{e}\,
\pd_{(\mu\bar{\sigma})}
\pd^{(\mu\bar{\sigma})}\bar{\phi}
+\frac{1}{4}\int d\bar{\sigma}
\,\bar{e}\,
\pd_{(\mu\bar{\sigma})}
\pd^{(\mu\bar{\sigma})}\,\bar{\psi}_{dd}
-\frac{1}{4}\int d\bar{\sigma}
\,\bar{e}\,
\pd_{(\mu\bar{\sigma})}
\pd^{(\mu\bar{\sigma})}\,
\int d\bar{\sigma}'\,
\bar{e}'\,
{\bar{\psi}^{(\mu'\bar{\sigma}')}}_{(\mu'\bar{\sigma}')}
=0,
}
around the flat 0-th order background (\ref{flatmetric}) under the static condition (\ref{static}) in the generalized harmonic gauge (\ref{nsh}). This becomes Laplace equation,
\Eq{
\int d\bar{\sigma}\,\bar{e}\,
\pd_{(\mu\bar{\sigma})}
\pd^{(\mu\bar{\sigma})}\bar{\phi}
=0, \label{eomscalar}
}
if the metric satisfies the Einstein equation (\ref{nspEin3}).

On the other hand, the equation of motion of the two-form field 
\begin{equation}
\bar{\nabla}_M(e^{-2\bar{\phi}} \bar{H}^{MNP})=0,
\end{equation}
is written as 
\begin{equation}
\hat{\nabla}_M \bar{H}^{MNP}=0,
\end{equation}
up to the first order in the fields, $\bar{h}_{IJ}$, $\bar{B}_{IJ}$, and $\bar{\phi}$. 
Furthermore, this becomes Laplace equation
\Eqr{
\int d\bar{\sigma}\,\bar{e}\,
\pd_{(\mu\bar{\sigma})}\pd^{(\mu\bar{\sigma})}
\,\bar{B}_{MN}=0,\label{eomB}
}
around the flat 0-th order background (\ref{flatmetric}) under the static condition (\ref{static}) in Lorentz gauge,
\begin{equation}
\hat{\nabla}_M \bar{B}^{MN}=0, 
\end{equation}
which is written as
\begin{equation}
\partial_{(\mu\bar{\sigma})} \bar{B}^{(\mu\bar{\sigma}) N}=0.
\label{BLorentz}
\end{equation}

We consider classical backgrounds corresponding to the string background configurations:
\begin{eqnarray}
\bar{\psi}_{dd}&=&0,  \\
\bar{\psi}_{d(\mu\bar{\sigma})}&=&0, \label{psidmu}\\
\bar{h}_{(\mu\bar{\sigma})(\nu\bar{\sigma}')}&=&
\frac{\bar{e}^3}{\sqrt{\bar{h}}}\,
g_{\mu\nu}(X(\bar{\sigma}))\delta_{\bar{\sigma}\bar{\sigma}'},  \label{psimunu}\\
\bar{B}_{d(\mu\bar{\sigma})}&=&0, \label{Bdmu} \\
\bar{B}_{(\mu\bar{\sigma})(\nu\bar{\sigma}')}&=&
\frac{\bar{e}^3}{\sqrt{\bar{h}}}\,
B_{\mu\nu}(X(\bar{\sigma}))\delta_{\bar{\sigma}\bar{\sigma}'},  \label{Bmunu}\\
\bar{\phi}&=& \int d \bar{\sigma} \bar{e} \Phi(X(\bar{\sigma})), \label{Phi}
\end{eqnarray}
where $g_{\mu\nu}(x)$ and $B_{\mu\nu}(x)$ satisfy gauge fixing conditions,
\begin{eqnarray}
\partial^{\mu} \psi_{\mu\nu}(x)&=&0, \nonumber \\
\partial^{\mu} B_{\mu\nu}(x)&=&0, \label{gaugefixing10d}
\end{eqnarray}
where
\Eqr{
\psi_{\mu \nu}=g_{\mu \nu}
-\frac{1}{2}\delta^{\alpha \beta}
g_{\alpha \beta}\delta_{\mu \nu}
+2 \delta_{\mu \nu}\Phi,
}
which imply (\ref{GaugeMetric}) and (\ref{BLorentz}).
Indeed, these are equivalent to
\begin{eqnarray}
\bar{G}_{dd}&=&-1,  \\
\bar{G}_{d(\mu\bar{\sigma})}&=&0, \\
\bar{G}_{(\mu\bar{\sigma})(\nu\bar{\sigma}')}&=&
\frac{\bar{e}^3}{\sqrt{\bar{h}}}\,
G_{\mu\nu}(X(\bar{\sigma}))\delta_{\bar{\sigma}\bar{\sigma}'}, \\
\bar{B}_{d(\mu\bar{\sigma})}&=&0,  \\
\bar{B}_{(\mu\bar{\sigma})(\nu\bar{\sigma}')}&=&
\frac{\bar{e}^3}{\sqrt{\bar{h}}}\,
B_{\mu\nu}(X(\bar{\sigma}))\delta_{\bar{\sigma}\bar{\sigma}'}, \\
\bar{\phi}&=& \int d \bar{\sigma} \bar{e} \Phi(X(\bar{\sigma})),
\end{eqnarray}
where
\Eq{
G_{\mu\nu}(x)=\delta_{\mu\nu}+g_{\mu\nu}(x) \,.
}
These are the string background configurations themselves \cite{Honda:2020sbl, Honda:2021rcd}.
If we impose that $g_{\mu\nu}(x)$, $B_{\mu\nu}(x)$ and  $\Phi(x)$  satisfy Laplace equations,
\begin{eqnarray}
\partial_{\rho} \partial^{\rho} g_{\mu\nu}(x)&=&0, \nonumber \\
\partial_{\rho} \partial^{\rho} B_{\mu\nu}(x)&=&0, \nonumber \\
\partial_{\rho} \partial^{\rho} \Phi(x)&=&0,  \label{Laplacesugra}
\end{eqnarray}
$\bar{G}_{MN}$, $\bar{B}_{MN}$ and $\bar{\phi}$ satisfy their equations of motion in string geometry theory\footnote{Under (\ref{eomscalar}),  
(\ref{nspEin3}) is equivalent to $\int d\bar{\sigma}\,\bar{e}\,
\pd_{(\mu\bar{\sigma})}\pd^{(\mu\bar{\sigma})}
\,\bar{h}_{MN}=0$, because (\ref{inverse}).},
(\ref{nspEin3}), (\ref{eomscalar}) and (\ref{eomB}), and
$G_{\mu \nu}$, $B_{\mu \nu}$ and $\Phi$ also satisfy their equations of motion  of the NS-NS sector in the supergravity.
Therefore, these string background configurations  in string geometry theory 
 represent perturbative string vacua parametrized by the on-shell fields in the supergravity as string backgrounds.

Next, we consider fluctuations around these vacua. The scalar fluctuation $\psi_{dd}$ represents the degrees of freedom of  perturbative strings in the case of the flat background as in \cite{Sato:2017qhj, Sato:2019cno, Sato:2020szq}. Thus, we also consider the scalar fluctuation $\psi_{dd}$ around the general perturbative vacua. We set the classical part of $\psi_{dd}$ as
\begin{equation}
\bar{\psi}_{dd}=\int {\cal D}X'(\bar{\tau}) G(X ; X')
\int d\bar{\sigma} \sqrt{\bar{h}}\,
\,\left[\alpha'R_{\bar{h}}
\Phi(X'(\bar{\sigma}))
+\frac{1}{\bar{e}^2}G_{\mu\nu}(X'(\bar{\sigma}))
\pd_{\bar{\sigma}} X^{'\mu} \pd_{\bar{\sigma}} X^{'\nu} 
\right],  \label{scalarclassicalfluctuation}
\end{equation}
where $R_{\bar{h}}$ is the scalar curvature of the two-dimensional metric $\bar{h}_{mn}$ and 
$G(X ; X')$ is a Green function on the flat string manifold given by 
\begin{equation}
G(X ; X')= \mathcal{N} \left[\int d\bar{\sigma}' 
\frac{\bar{e}^{'2}}{\sqrt{\bar{h}'}}\left(X^\mu(\bar{\sigma}')
-{X'}^\mu(\bar{\sigma}')\right)^2
\right]^{\frac{2-D}{2}}, \label{GreenFunc}
\end{equation}
which satisfies 
\begin{equation}
\int d\bar{\sigma}\sqrt{\bar{h}}
\frac{1}{\bar{e}}\frac{\pd}{\pd X^\mu(\bar{\sigma})}
\frac{1}{\bar{e}}\frac{\pd}{\pd X_\mu(\bar{\sigma})}
G(X ; X')=\delta(X-X'), \label{GreenEq}
\end{equation}
where $\mathcal{N}$ is a normalizing constant.
A derivation is given in the appendix.  
As a result, $\bar{\psi}_{dd}$ is not on-shell but satisfies
\Eqr{
&&\int d\bar{\sigma}\sqrt{\bar{h}}
\frac{1}{\bar{e}}\frac{\pd}{\pd X^\mu(\bar{\sigma})}
\frac{1}{\bar{e}}\frac{\pd}{\pd X_\mu(\bar{\sigma})}
\,\bar{\psi}_{dd}\nn\\
&&~~~~~
=
\int d\bar{\sigma}\sqrt{\bar{h}}\left[\alpha'\,
R_{\bar{h}}
\Phi(X(\bar{\sigma}))+\frac{1}{\bar{e}^2}
G_{\mu\nu}(X(\bar{\sigma}))
\pd_{\bar{\sigma}} X^\mu \pd_{\bar{\sigma}} X^\nu
\right]. \label{NoOnshell}
}

Furthermore, we consider the quantum part of $\psi_{dd}$,
\begin{equation}
\tilde{\psi}_{dd}=\frac{D-1}{D-2}\tilde{\phi},
\end{equation}
where $\frac{D-1}{D-2}$ is introduced for later convenience. 
Totally, 
\Eqr{
G_{MN}=\hat{G}_{MN}+\bar{h}_{MN}+\tilde{G}_{MN}, \label{G}
}
where $\hat{G}_{MN}$  is  given by (\ref{flatmetric}), $\bar{h}_{MN}$  is given by (\ref{compo}) with (\ref{scalarclassicalfluctuation}), (\ref{psidmu}), (\ref{psimunu}) and (\ref{Phi}), and $\tilde{G}_{MN}$ is given by
\Eqr{
&&\hspace{-1cm}
\tilde{G}_{dd}=\tilde{\phi},~~~~
\tilde{G}_{d(\mu\bar{\sigma})}=0,~~~~
\tilde{G}_{(\mu\bar{\sigma})(\mu'\bar{\sigma}')}
=\frac{1}{D-2}
\frac{\bar{e}^3}{\sqrt{\bar{h}}}\,
\tilde{\phi}\,\delta_{(\mu\bar{\sigma})(\mu'\bar{\sigma}')}\,.
  \label{gh}
}

%
%
\section{Deriving the path integrals of the perturbative strings on curved backgrounds
}
In this section, we will derive the path integrals of the perturbative strings up to any order from the tree-level two-point correlation functions of the quantum scalar fluctuations of the metric. In order to obtain a propagator, we add a gauge fixing term corresponding to 
(\ref{nsh}) into the action (\ref{action}) and obtain
\Eqr{
S&=&
\int {\cal D}\bar{\tau}\, {\cal D}\bar{h}\,{\cal D}X(\bar{\tau})
\sqrt{-G}\,e^{-2\phi}
\left[R+4\nabla_I\phi\nabla^I\phi
-\frac{1}{2}|H|^2\right.\nn\\
&&\left.
-\frac{1}{2}\left\{\bar{\nabla}^N\left(\tilde{G}_{MN}-
\frac{1}{2} \bar{G}^{IJ} \tilde{G}_{IJ}\bar{G}_{MN}
+2\bar{G}_{MN}\,\bar{\phi}\right)\right\}^2
\right],
  \label{s-action}
}
where we abbreviate the  Faddeev-Popov ghost term because it does not contribute to the tree-level two-point correlation functions of the metrics. 
By substituting Eqs.~(\ref{G}), (\ref{Bdmu}),  
(\ref{Bmunu}) and (\ref{Phi})  that do not necessarily satisfy the equations of motion (\ref{Laplacesugra}) into (\ref{s-action}),  
This is expressed as 
\Eqr{
S=\int 
 {\cal D}\bar{\tau}\, {\cal D}\bar{h}\,{\cal D}X(\bar{\tau})
\,
\left(c_0+c_1\,\tilde{\phi}
+\tilde{\phi}\,c_2\,\tilde{\phi}
+\tilde{\phi}\, \int d\bar{\sigma}\,\bar{e}\,
\int d\bar{\sigma}'\,\bar{e}'c^{(\mu\bar{\sigma})(\mu'\bar{\sigma}')}
\,\pd_{(\mu\bar{\sigma})}\pd_{(\mu'\bar{\sigma}')}\tilde{\phi}
\right), \label{fixedaction}
}
where 
\Eqrsubl{c}{
&&\hspace{-1cm}c_0=
-\frac{1}{D-1}\int d\bar{\sigma}\,\bar{e}\,
\pd_{(\mu\bar{\sigma})}\pd^{(\mu\bar{\sigma})}\,\bar{\psi}_{dd}
-\frac{4D}{D-1}\int d\bar{\sigma}\,\bar{e}\,
\pd_{(\mu\bar{\sigma})}
\pd^{(\mu\bar{\sigma})}\bar{\phi}\nn\\
&&~~~~~+\frac{1}{D-1}\int d\bar{\sigma}\,\bar{e}\,
\pd_{(\mu\bar{\sigma})}\pd^{(\mu\bar{\sigma})}
\int d\bar{\sigma}'\,\bar{e}'\,
{\bar{\psi}_{(\mu'\bar{\sigma}')}}^{(\mu'\bar{\sigma}')},\\
&&\hspace{-1cm}c_1=
\frac{1}{2}\int d\bar{\sigma}\,
\bar{e}\,
\pd_{(\mu\bar{\sigma})}
\pd^{(\mu\bar{\sigma})}\,\bar{\psi}_{dd}\nn\\
&&
+\frac{1}{2(D-2)}\int d\bar{\sigma}\,\bar{e}\,
\pd_{(\mu\bar{\sigma})}\pd^{(\mu\bar{\sigma})}
\int d\bar{\sigma}'\,\bar{e}'\,
{\bar{\psi}_{(\mu'\bar{\sigma}')}}^{(\mu'\bar{\sigma}')},\\
&&\hspace{-1cm}c_2=
\frac{1}{4}\int d\bar{\sigma}\,\bar{e}\,
\pd_{(\mu\bar{\sigma})}
\pd^{(\mu\bar{\sigma})}\,\bar{\psi}_{dd}\nn\\
&&
-\frac{1}{4(D-2)^2}\int d\bar{\sigma}\,\bar{e}\,
\pd_{(\mu\bar{\sigma})}\pd^{(\mu\bar{\sigma})}\,
\int d\bar{\sigma}'\,\bar{e}'\,
{\bar{\psi}_{(\mu'\bar{\sigma}')}}^{(\mu'\bar{\sigma}')},\\
&&\hspace{-1cm}
c^{(\mu\bar{\sigma})(\mu'\bar{\sigma}')}=
\left[
\frac{D-1}{4(D-2)}
+\frac{1}{2}
\,\bar{\psi}_{dd}
+\frac{1}{2(D-2)}\int d\bar{\sigma}''\,\bar{e}''\,
{\bar{\psi}_{(\mu''\bar{\sigma}'')}}^{(\mu''\bar{\sigma}'')}
\right.\nn\\
&&\left.~~~~~
-\frac{2}{D-2}
\bar{\phi}
\right]
\delta^{(\mu\bar\sigma)(\mu'\bar{\sigma}')}
-\frac{D-1}{4(D-2)}
\bar{\psi}^{(\mu\bar{\sigma})(\mu'\bar{\sigma}')},
}
up to the first order in the classical fields and the second order in $\tilde{\phi}$.
Here, we take the regularization parameter  $D\rightarrow\infty$. Then, (\ref{fixedaction}) becomes
\Eqr{
\hspace{-0.5cm}S&=&\int 
 {\cal D}\bar{\tau}\, {\cal D}\bar{h}\,{\cal D}X(\bar{\tau})
\,
\left[-4\int d\bar{\sigma}\,\bar{e}\,\pd_{(\mu\bar{\sigma})}
\pd^{(\mu\bar{\sigma})}\bar{\phi}
+\frac{1}{2}\,\int d\bar{\sigma}\,\bar{e}\,
\pd_{(\mu\bar{\sigma})}\pd^{(\mu\bar{\sigma})}\bar{\psi}_{dd}
\,\tilde{\phi}
\right.\nn\\
&&\hspace{-0.7cm}
+\,\tilde{\phi}\,\frac{1}{4}\int d\bar{\sigma}\,\bar{e}\,
\pd_{(\mu\bar{\sigma})}\pd^{(\mu\bar{\sigma})}\bar{\psi}_{dd}
\,\tilde{\phi}
+\tilde{\phi}\,
\left(\frac{1}{4}+\frac{1}{2}\bar{\psi}_{dd}\right)\int d\bar{\sigma}\,\bar{e}\,
\,\pd_{(\mu\bar{\sigma})}
\pd^{(\mu\bar{\sigma})}\tilde{\phi}
%
%
\nn\\
&&\left.\hspace{-0.7cm}
-\frac{1}{4}\tilde{\phi}\,\int d\bar{\sigma}\,\bar{e}\,\int d\bar{\sigma}'\,\bar{e}'\,
\bar{\psi}^{(\mu\bar{\sigma})(\mu'\bar{\sigma}')}
\,\pd_{(\mu\bar{\sigma})}
\pd_{(\mu'\bar{\sigma}')}\tilde{\phi}
\right].
  \label{action2}
}
By shifting the field $\tilde{\phi}$ as $\tilde{\phi}=\tilde{\phi}'
-\frac{2}{3}$, the first order term in $\tilde{\phi}'$ vanishes as
\Eqr{
\hspace{-0.5cm}S
&=&\int 
 {\cal D}\bar{\tau}\, {\cal D}\bar{h}\,{\cal D}X(\bar{\tau})
\biggl[\tilde{\phi}'\,\frac{1}{4}\int d\bar{\sigma}\,\bar{e}\,
\pd_{(\mu\bar{\sigma})}\pd^{(\mu\bar{\sigma})}\bar{\psi}_{dd}
\,\tilde{\phi}'
+\tilde{\phi}'\,
\left(\frac{1}{4}+\frac{1}{2}\bar{\psi}_{dd}+\frac{1}{8}\hat{G}^{IJ}
\bar{h}_{IJ}\right)\int d\bar{\sigma}\,\bar{e}\,
\,\pd_{(\mu\bar{\sigma})}
\pd^{(\mu\bar{\sigma})}\tilde{\phi}'
%
%
\nn\\
&&\left.\hspace{-0.7cm}
-\frac{1}{4}\tilde{\phi}'\,\int d\bar{\sigma}\,\bar{e}\,\int d\bar{\sigma}'\,\bar{e}'\,
\bar{h}^{(\mu\bar{\sigma})(\mu'\bar{\sigma}')}
\,\pd_{(\mu\bar{\sigma})}
\pd_{(\mu'\bar{\sigma}')}\tilde{\phi}'
\right],
  \label{action-1st-2}
}
where surface terms are dropped and the gauge fixing condition in (\ref{gaugefixing10d}) and a relation (\ref{spsi}) are applied. 
By normalizing the leading part of the kinetic term as $\tilde{\phi}'=2(1-\bar{\psi}_{dd}-\frac{1}{4}\hat{G}^{IJ}
\bar{h}_{IJ})\tilde{\phi}''$, 
we have 
\Eqr{
S&=&\int 
 {\cal D}\bar{\tau}\, {\cal D}\bar{h}\,{\cal D}X(\bar{\tau})
\left[\int d\bar{\sigma}\,\bar{e}\,
\,\pd_{(\mu\bar{\sigma})}\pd^{(\mu\bar{\sigma})}\bar{\psi}_{dd}\,
\left(\tilde{\phi}''\right)^2
+
\tilde{\phi}''\,\int d\bar{\sigma}\,\bar{e}\,
\pd_{(\mu\bar{\sigma})}\pd^{(\mu\bar{\sigma})}\tilde{\phi}''
\right.
\nn\\
&&~~~\left.
-\tilde{\phi}''\int d\bar{\sigma}\,\bar{e}\,\int d\bar{\sigma}'\bar{e}'\,
\,\bar{h}_{(\mu\bar{\sigma})(\mu'\bar{\sigma}')}\,
\pd^{(\mu\bar{\sigma})}\pd^{(\mu'\bar{\sigma}')}
\tilde{\phi}''
\right].
  \label{action-1st-3}
}
This can be written as
\Eqr{
S=-2
\int {\cal D}\bar{\tau}\,{\cal D}\bar{h}\,{\cal D}X(\bar{\tau})\,\tilde{\phi}''\,
H\left(-i\frac{1}{\bar{e}}\frac{\pd}{\pd X},~
X,~\bar{h}\right)
\tilde{\phi}'',
}
where
\Eqr{
&&H(p_X,~X,~\bar{h})
=\frac{1}{2}\int d\bar{\sigma}\sqrt{\bar{h}}\left(
p^\mu_X\right)^2
-\frac{1}{2}\int d\bar{\sigma}\sqrt{\bar{h}}
g_{\mu\nu}(X(\bar{\sigma}))
p^\mu_X
p^{\nu}_X\nn\\
&&~~~~~~~~~~~
\qquad \qquad -\frac{1}{2}\int d\bar{\sigma}
\sqrt{\bar{h}}
\frac{1}{\bar{e}}\frac{\pd}{\pd X^\mu}
\frac{1}{\bar{e}}\frac{\pd}{\pd X_\mu}
\,\bar{\psi}_{dd}\, \nn\\
&&~~~~~~
\qquad \qquad \qquad +\int d\bar{\sigma} \bar{n}^{\bar{\sigma}} \pd_{\bar{\sigma}} X^\mu 
\bar{e}p_{\mu\,X}
+\int d\bar{\sigma} 
\,i\,\frac{\sqrt{\bar{h}}}{\bar{e}^2}\,
\pd_{\bar{\sigma}} X^\nu {B_{\nu}}^\mu
\bar{e}p_{\mu\,X}.
}
Here we have added terms 
\Eqr{
0=-2
\int {\cal D}\bar{\tau}\,{\cal D}\bar{h}\,{\cal D}X(\bar{\tau})\,\tilde{\phi}''\,
\left(-i\int d\bar{\sigma} \bar{n}^{\bar{\sigma}} \pd_{\bar{\sigma}} X^\mu
\frac{\pd}{\pd X^\mu}
+\int d\bar{\sigma} \frac{\sqrt{\bar{h}}}{\bar{e}^2}\,
\pd_{\bar{\sigma}} X^\nu 
{B_{\nu}}^\mu\frac{\pd}{\pd X^\mu},\right)
\tilde{\phi}'',
}
which is true because of the gauge fixing condition (\ref{gaugefixing10d}).

The propagator for $\tilde{\phi}$ defined by
\begin{equation}
\Delta_F(\bar{h}, X(\bar{\tau}); \; \bar{h},'  X'(\bar{\tau}'))=<\tilde{\phi} (\bar{h}, X(\bar{\tau}))
\tilde{\phi}(\bar{h},'  X'(\bar{\tau}'))>
\end{equation}
satisfies
\begin{equation}
H(-i\frac{1}{\bar{e}}\frac{\partial}{\partial X(\bar{\tau})}, X(\bar{\tau}), \bar{h})
\Delta_F(\bar{h}, X(\bar{\tau}); \; \bar{h},'  X'(\bar{\tau}'))
=
\delta(\bar{h}-\bar{h}') \delta(X(\bar{\tau})-X'(\bar{\tau}')). \label{delta}
\end{equation}
In order to obtain a Schwinger representation of the propagator, we use the operator formalism $(\hat{\bar{h}}, \hat{X}(\bar{\tau}))$ of the first quantization, whereas the conjugate momentum is written as $(\hat{p}_{\bar{h}},  \hat{p}_{X}(\bar{\tau}))$. The eigen state is given by $|\bar{h}, X(\bar{\tau})>$. 

Since \eqref{delta} means that $\Delta_F$ is an inverse of $H$, $\Delta_F$ can be expressed by a matrix element of the operator $\hat{H}^{-1}$ as
\begin{equation}
\Delta_F(\bar{h}, X(\bar{\tau}); \; \bar{h},'  X'(\bar{\tau}'))
=
<\bar{h}, X(\bar{\tau})| \hat{H}^{-1}(\hat{p}_{X}(\bar{\tau}), \hat{X}(\bar{\tau}), \hat{\bar{h}}) |\bar{h},' X'(\bar{\tau}') >.
\label{InverseH}
\end{equation}
On the other hand,
\begin{eqnarray}
\hat{H}^{-1}=  i \int _0^{\infty} dT e^{-iT\hat{H}}, \label{IntegralFormula}
\end{eqnarray}
because
\begin{equation}
\lim_{\epsilon \to 0+} \int _0^{\infty} dT e^{-T(i\hat{H}+\epsilon)}
=
\lim_{\epsilon \to 0+}  \left[\frac{1}{-(i\hat{H}+\epsilon)} e^{-T(i\hat{H}+\epsilon)}
\right]_0^{\infty}
=-i\hat{H}^{-1}.
\end{equation}
This fact and \eqref{InverseH} imply
\begin{equation}
\Delta_F(\bar{h}, X(\bar{\tau}); \; \bar{h},'  X'(\bar{\tau}'))
=
i \int _0^{\infty} dT <\bar{h}, X(\bar{\tau})|  e^{-iT\hat{H}} |\bar{h},' X'(\bar{\tau}') >.
\end{equation}
In order to define two-point correlation functions that is invariant under the general coordinate transformations in the string geometry, we define in and out states as
\begin{align}
||X_i \,|\,h_f, ; h_i>_{in} :=& \int_{h_i}^{h_f} \mathcal{D}h'|\bar{h},' X_i:=X'(\bar{\tau}'=-\infty)> 
\nonumber \\
<X_f\,|\,h_f, ; h_i||_{out}  :=& \int_{h_i}^{h_f} \mathcal{D} h <\bar{h}, X_f:=X(\bar{\tau}=\infty)|,
\end{align}
where $h_i$ and $h_f$ represent the metrics of the cylinders at $\bar{\tau}=\pm \infty$, respectively. $\int$ in $\int \mathcal{D} h$ includes $\sum_{\mbox{compact topologies}}$, where $\mathcal{D}h$ is the invariant measure\footnote{The invariant measure is defined implicitly by the most general invariant norm without derivatives for elements $\delta h_{mn}$ of the tangent space of the metric, $||\delta h||^2=\int d^2 \sigma \sqrt{h}(h^{mp} h^{nq}+C h^{mn}h^{pq} )\delta h_{mn} \delta h_{pq}$ with $C$ an arbitrary constant, and a normalization $\int \mathcal{D} \delta h \exp^{-\frac{1}{2}||\delta h||^2}=1.$}. of the metrics $h_{mn}$ on the two-dimensional Riemannian manifolds $\Sigma$. $h_{mn}$ and $\bar{h}_{mn}$ are related to each others by the diffeomorphism and the Weyl transformations. When we insert asymptotic states, we integrate out $X_f$, $X_i$, $h_f$ and $h_i$ in the two-point correlation function for these states;
\begin{eqnarray}
\Delta_F(X_f; X_i|h_f, ; h_i) := i \int _0^{\infty} dT <X_f \,|\,h_f, ; h_i||_{out}  e^{-iT\hat{H}} ||X_i \,|\,h_f, ; h_i>_{in}.
\end{eqnarray}

This can be written as in \cite{Sato:2017qhj}\footnote{The correlation function is zero if $h_i$ and $h_f$ of the in state do not coincide with those of the out states, because of the delta functions in the sixth line.},
\begin{eqnarray}
&&\Delta_F(X_f; X_i|h_f, ; h_i) \nonumber \\
&:=&i\int _0^{\infty} dT <X_f \,|\,h_f, ; h_i||_{out}  e^{-iT\hat{H}} ||X_i \,|\,h_f, ; h_i>_{in}\nonumber \\
&=&i\int _0^{\infty} dT  \lim_{N \to \infty} \int_{h_i}^{h_f} \mathcal{D} h \int_{h_i}^{h_f} \mathcal{D} h' 
\prod_{n=1}^N \int d \bar{h}_{n} dX_n(\bar{\tau}_n) 
\nonumber \\
&&\prod_{m=0}^N <\bar{h}_{m+1}, X_{m+1}(\bar{\tau}_{m+1})| e^{-i\frac{1}{N}T \hat{H}} |\bar{h}_{m}, X_{m}(\bar{\tau}_m),> \nonumber \\
&=&i\int _0^{\infty} dT_0 \lim_{N \to \infty} \int d T_{N+1} \int_{h_i}^{h_f} \mathcal{D} h \int_{h_i}^{h_f} \mathcal{D} h' \prod_{n=1}^N \int d T_n d \bar{h}_{n} dX_n(\bar{\tau}_n)  \nonumber \\
&&\prod_{m=0}^N  < X_{m+1}(\bar{\tau}_{m+1})| e^{-i\frac{1}{N}T_m \hat{H}} |X_{m}(\bar{\tau}_m)>\delta(\bar{h}_{m}-\bar{h}_{m+1})\delta(T_{m}-T_{m+1}) \nonumber \\
&=&i\int _0^{\infty} dT_0 \lim_{N \to \infty} d T_{N+1} \int_{h_i}^{h_f} \mathcal{D} h \prod_{n=1}^N \int d T_n  dX_n(\bar{\tau}_n)    \prod_{m=0}^N \int dp_{T_m}  dp_{X_m}(\bar{\tau}_m)  \nonumber \\
&&\exp \Biggl(i\sum_{m=0}^N \Delta t
\Bigl(p_{T_m} \frac{T_{m}-T_{m+1}}{\Delta t} 
+p_{X_m}(\bar{\tau}_m)\cdot \frac{X_{m}(\bar{\tau}_m)-X_{m+1}(\bar{\tau}_{m+1})}{\Delta t} 
\nonumber \\
&&-T_m H( p_{X_m}(\bar{\tau}_m), X_{m}(\bar{\tau}_m), \bar{h})\Bigr) \Biggr) \nonumber \\
&=&
i \int_{h_i X_i}^{h_f, X_f}  \mathcal{D} h \mathcal{D} X(\bar{\tau}) 
\int \mathcal{D} T  
\int 
\mathcal{D} p_T
\mathcal{D}p_{X} (\bar{\tau})
 \nonumber \\
&&
\exp \Biggl(i\int_{-\infty}^{\infty} dt \Bigr(
 p_{T}(t) \frac{d}{dt} T(t) 
+ p_{X}(\bar{\tau}(t), t)\cdot \frac{d}{dt} X(\bar{\tau}(t), t)\nonumber \\
&&
-T(t) H( p_{X}(\bar{\tau}(t), t), X(\bar{\tau}(t), t), \bar{h})\Bigr) \Biggr),  \label{canonicalform}
\end{eqnarray}
where $p_{X}(\bar{\tau}(t), t) \cdot \frac{d}{dt} X(\bar{\tau}(t), t):= \int d\bar{\sigma} \bar{e} p_{X}^{\mu}(\bar{\tau}(t), t) \frac{d}{dt} X_{\mu}(\bar{\tau}(t), t)$.
 $\bar{h}_{ 0}=\bar{h}'$, $X_{0}(\bar{\tau}_0)=X_i$, $\bar{\tau}_0=-\infty$, $\bar{h}_{ N+1}=\bar{h}$, $X_{N+1}(\bar{\tau}_{N+1})=X_f$, $\bar{\tau}_{N+1}=\infty$, and $\Delta t:=\frac{1}{\sqrt{N}}$.
A trajectory of points $[\bar{\Sigma}, X(\bar{\tau})]$ is necessarily continuous in $\mathcal{M}_D$ so that the kernel $<\bar{h}_{m+1}, X_{m+1}(\bar{\tau}_{m+1})| e^{-i\frac{1}{N}T_m \hat{H}} |\bar{h}_{m}, X_{m}(\bar{\tau}_m)>$ in the fourth line is non-zero when $N \to \infty$.

By integrating out $p_{X}(\bar{\tau}(t), t)$, we move from the canonical formalism to the Lagrange formalism.
Because the exponent of (\ref{canonicalform}) is at most the second order in $p_{X}(\bar{\tau}(t), t)$,
integrating out $p_{X}(\bar{\tau}(t), t)$ is equivalent to 
substituting into (\ref{canonicalform}), the solution $p_{X}(\bar{\tau}(t), t)$ of  
\Eqr{
-i\bar{e}\frac{d}{dt}X^\mu
+iT\bar{e}\left( 
\bar{n}^{\bar{\sigma}}\pd_{\bar{\sigma}} X^\mu
+ i\,
\frac{\sqrt{\bar{h}}}
{\bar{e}^2}\,
\pd_{\bar{\sigma}} X^\nu {B_{\nu}}^\mu\right)
+iT\, \sqrt{\bar{h}} p^\mu_X 
-iT\sqrt{\bar{h}}
g^{\mu\nu}(X)
p_{\nu X}=0,
  \label{dS'}
}
which is obtained by differentiating  the exponent of (\ref{canonicalform})
with respect to $p_{X}(\bar{\tau}(t), t)$. The solution is given by  
\Eqr{
p_{\mu X}&=&
\left[\delta_{\mu\nu}
+g_{\mu\nu}(X)
\right]
\frac{1}{T}\frac{\bar{e}}
{\sqrt{\bar{h}}}
\left[\frac{d}{dt}X^\nu
-T\left(\bar{n}^{\bar{\sigma}}\pd_{\bar{\sigma}} X^\nu
+\,i
\,\frac{\sqrt{\bar{h}}}{\bar{e}^2}\,
\pd_{\bar{\sigma}} X^\gamma 
{B_{\gamma}}^\nu (X) \right)
\right],
  \label{pmu}
}
up to the first order in the classical backgrounds $g_{\mu\nu}(X)$ and $B_{\mu \nu}(X)$.
By substituting this, we obtain
\begin{align}
&\Delta_F(X_f; X_i|h_f ; h_i) \nonumber \\
&\quad
=i \int_{h_i X_i}^{h_f, X_f} 
\mathcal{D} T
\mathcal{D} h  \mathcal{D} X(\bar{\tau})
\mathcal{D} p_T 
\nonumber \\
&\qquad\quad
\exp \Biggl(i \int_{-\infty}^{\infty} dt \Bigl( p_{T}(t) \frac{d}{dt} T(t)   \nonumber \\
&\qquad\qquad\qquad
+\int d\bar{\sigma} \sqrt{\bar{h}}G_{\mu\nu}(X(\bar{\tau}(t), t)) ( 
\frac{1}{2}\bar{h}^{00}\frac{1}{T(t)}\partial_{t} X^{\mu}(\bar{\tau}(t), t)\partial_{t} X^{\nu}(\bar{\tau}(t), t) \nonumber \\
&\qquad\qquad\qquad
 +\bar{h}^{01}\partial_{t} X^{\mu}(\bar{\tau}(t), t)\partial_{\bar{\sigma}} X^{\nu}(\bar{\tau}(t), t) 
+\frac{1}{2}\bar{h}^{11}T(t)\partial_{\bar{\sigma}} X^{\mu}(\bar{\tau}(t), t)\partial_{\bar{\sigma}} X^{\nu}(\bar{\tau}(t), t)
)  \nonumber \\
&\qquad\qquad\qquad+\int d\bar{\sigma}\,i\,
B_{\mu\nu} (X(\bar{\tau}(t), t))
\partial_{t} X^{\mu}(\bar{\tau}(t), t)\partial_{\bar{\sigma}} X^{\nu}(\bar{\tau}(t), t) 
\nonumber \\
&\qquad\qquad\qquad
+\frac{1}{2}\int d\bar{\sigma}\sqrt{\bar{h}}\,T(t)\,
\alpha'\,R_{\bar{h}}
\Phi (X(\bar{\tau}(t), t))
\Bigr) \Biggr), \label{pathint1}
\end{align}
where we use (\ref{NoOnshell}) and the ADM decomposition of the two-dimensional metric,
\Eqr{
\bar{h}_{mn}=
\left(
\begin{array}{cc}
\bar{n}^2+ \bar{n}_{\bar{\sigma}} \bar{n}^{\bar{\sigma}} & \bar{n}_{\bar{\sigma}} \\
\bar{n}_{\bar{\sigma}} & \bar{e}^2
\end{array}
\right),
~~~~~~
\sqrt{\bar{h}}=\bar{n}\bar{e},
~~~~~~\bar{h}^{mn}=
\begin{pmatrix}
\frac{1}{\bar{n}^2} & -\frac{\bar{n}^{\bar{\sigma}}}
{\bar{n}^2} \\
-\frac{\bar{n}^{\bar{\sigma}}}{\bar{n}^2} &
\bar{e}^{-2}+\left(
\frac{\bar{n}^{\bar{\sigma}}}{\bar{n}}\right)^2
\end{pmatrix}\,.
}
In this way, the Green function can generate all the terms without $\bar{\tau}$ derivatives in the string action as in (\ref{NoOnshell}), but cannot do those with $\bar{\tau}$ derivatives,   which need  to be derived non-trivially, because the coordinates  $ X^{\mu}(\bar{\tau})$ in string geometry theory are defined on the $\bar{\tau}$ constant lines.  
We should note that the time derivative in \eqref{pathint1} is in terms of $t$, not $\bar{\tau}$ at this moment. In the following, we will see that $t$ can be fixed to $\bar{\tau}$ by using a reparametrization of $t$ that parametrizes a trajectory.

By inserting
$\int \mathcal{D}c \mathcal{D}b
e^{\int_0^{1} dt \left(\frac{d b(t)}{dt} \frac{d c(t)}{dt}\right)
},$
where $b(t)$ and $c(t)$ are $bc$-ghost, we obtain 
\begin{align}
&\Delta_F(X_f; X_i|h_f ; h_i) \nonumber \\
&\quad
=Z_0 \int_{h_i X_i}^{h_f, X_f} 
\mathcal{D} T
\mathcal{D} h  \mathcal{D} X(\bar{\tau})
\mathcal{D} p_T
\mathcal{D}c \mathcal{D}b  \nonumber \\
&\qquad\quad
\exp \Biggl(- \int_{-\infty}^{\infty} dt \Bigl(
-i p_{T}(t) \frac{d}{dt} T(t)  
 +\frac{d b(t)}{dt} \frac{d (T(t) c(t))}{dt}\nonumber \\
&\qquad\qquad\qquad
+\int d\bar{\sigma} \sqrt{\bar{h}}G_{\mu\nu}(X(\bar{\tau}(t), t))  ( 
\frac{1}{2}\bar{h}^{00}\frac{1}{T(t)}\partial_{t} X^{\mu}(\bar{\tau}(t), t)\partial_{t} X^{\nu}(\bar{\tau}(t), t) \nonumber \\
&\qquad\qquad\qquad
+\bar{h}^{01}\partial_{t} X^{\mu}(\bar{\tau}(t), t)\partial_{\bar{\sigma}} X^{\nu}(\bar{\tau}(t), t) 
+\frac{1}{2}\bar{h}^{11}T(t)\partial_{\bar{\sigma}} X^{\mu}(\bar{\tau}(t), t)\partial_{\bar{\sigma}} X^{\nu}(\bar{\tau}(t), t)
)  \nonumber \\
&\qquad\qquad\qquad+\int d\bar{\sigma}\,i\,
B_{\mu\nu} (X(\bar{\tau}(t), t))
\partial_{t} X^{\mu}(\bar{\tau}(t), t)\partial_{\bar{\sigma}} X^{\nu}(\bar{\tau}(t), t) 
\nonumber \\
&\qquad\qquad\qquad
+\frac{1}{2}\int d\bar{\sigma}\sqrt{\bar{h}}\,T(t)\,
\alpha'\,R_{\bar{h}}
\Phi (X(\bar{\tau}(t), t))\Bigr) \Biggr), 
\label{PropWMult}
\end{align}
where we redefine as $c(t) \to T(t) c(t)$, and $Z_0$ represents an overall constant factor. In the following, we will rename it $Z_1, Z_2, \cdots$ when the factor changes. The integrand variable $p_T (t)$ plays the role of the Lagrange multiplier providing the following condition,
\begin{align}
F_1(t):=\frac{d}{dt}T(t)=0,
\label{F1gauge}
\end{align}
which can be understood as a gauge fixing condition. Indeed, by choosing this gauge in
\begin{align}
&\Delta_F(X_f; X_i|h_f ; h_i) \nonumber \\
&\quad
=Z_1 \int_{h_i X_i}^{h_f, X_f} 
\mathcal{D} T
\mathcal{D} h  \mathcal{D} X(\bar{\tau})
\nonumber \\
&\qquad\quad
\exp \Biggl(- \int_{-\infty}^{\infty} dt \Bigl(
\int d\bar{\sigma} \sqrt{\bar{h}} G_{\mu\nu}(X(\bar{\tau}(t), t)) ( 
\frac{1}{2}\bar{h}^{00}\frac{1}{T(t)}\partial_{t} X^{\mu}(\bar{\tau}(t), t)\partial_{t} X^{\nu}(\bar{\tau}(t), t)
\nonumber \\
&\qquad\qquad\qquad
 +\bar{h}^{01}\partial_{t} X^{\mu}(\bar{\tau}(t), t)\partial_{\bar{\sigma}} X^{\nu}(\bar{\tau}(t), t) 
+\frac{1}{2}\bar{h}^{11}T(t)\partial_{\bar{\sigma}} X^{\mu}(\bar{\tau}(t), t)\partial_{\bar{\sigma}} X^{\nu}(\bar{\tau}(t), t)
)  \nonumber \\
&\qquad\qquad\qquad+\int d\bar{\sigma}\,i\,
B_{\mu\nu} (X(\bar{\tau}(t), t))
\partial_{t} X^{\mu}(\bar{\tau}(t), t)\partial_{\bar{\sigma}} X^{\nu}(\bar{\tau}(t), t) 
\nonumber \\
&\qquad\qquad\qquad
+\frac{1}{2}\int d\bar{\sigma}\sqrt{\bar{h}}\,T(t)\,
\alpha'\,R_{\bar{h}}
\Phi (X(\bar{\tau}(t), t))\Bigr) \Biggr),
\label{pathint2}
\end{align}
we obtain \eqref{PropWMult}.
The expression \eqref{pathint2} has a manifest one-dimensional diffeomorphism symmetry with respect to $t$, where $T(t)$ is transformed as an einbein \cite{Schwinger0}. 

Under $\frac{d\bar{\tau}}{d\bar{\tau}'}=T(t)$, which implies
\begin{eqnarray}
\bar{h}^{00}&=&T^2\bar{h}^{'00}, \nonumber \\
\bar{h}^{01}&=&T\bar{h}^{'01}, \nonumber \\
\bar{h}^{11}&=&\bar{h}^{'11}, \nonumber \\
\sqrt{\bar{h}}&=&\frac{1}{T}\sqrt{\bar{h}'}, \nonumber \\
X^{\mu}(\bar{\tau}(t), t)&=&X^{'\mu}(\bar{\tau}'(t), t).
\end{eqnarray}

$T(t)$ disappears in \eqref{pathint2} and we obtain 
\begin{align}
&\Delta_F(X_f; X_i|h_f ; h_i) \nonumber \\
&=
Z_2 \int_{h_i X_i}^{h_f, X_f} 
\mathcal{D} h  \mathcal{D} X(\bar{\tau})
\nonumber \\
&\qquad
\exp \Biggl(- \int_{-\infty}^{\infty} dt \Bigl(
\int d\bar{\sigma} \sqrt{\bar{h}}G_{\mu\nu}(X(\bar{\tau}(t), t))  ( 
\frac{1}{2}\bar{h}^{00}\partial_{t} X^{\mu}(\bar{\tau}(t), t)\partial_{t} X^{\nu}(\bar{\tau}(t), t) 
\nonumber \\ &+\bar{h}^{01}\partial_{t} X^{\mu}(\bar{\tau}(t), t)\partial_{\bar{\sigma}} X^{\nu}(\bar{\tau}(t), t) 
+\frac{1}{2}\bar{h}^{11}\partial_{\bar{\sigma}} X^{\mu}(\bar{\tau}(t), t)\partial_{\bar{\sigma}} X^{\nu}(\bar{\tau}(t), t)
)  \nonumber \\
&+\int d\bar{\sigma}\,i\,
B_{\mu\nu} (X(\bar{\tau}(t), t))
\partial_{t} X^{\mu}(\bar{\tau}(t), t)\partial_{\bar{\sigma}} X^{\nu}(\bar{\tau}(t), t) 
\nonumber \\
&
+\frac{1}{2}\int d\bar{\sigma}\sqrt{\bar{h}}\,
\alpha'\,R_{\bar{h}}
\Phi (X(\bar{\tau}(t), t))\Bigr) \Biggr). \label{pathint3}
\end{align}
This action is still invariant under the diffeomorphism with respect to t if $\bar{\tau}$ transforms in the same way as $t$. 

\begin{figure}[htb]
\centering
\includegraphics[width=6cm]{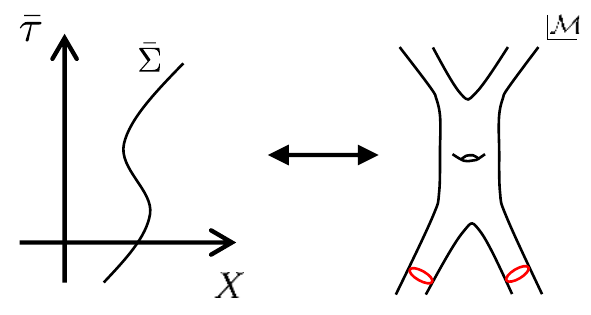}
\caption{A path and a Riemann surface. The line on the left is a trajectory in the path integral. The trajectory parametrized by $\bar{\tau}$ from $-\infty$ to $\infty$, represents a Riemann surface with fixed punctures in $\mathcal{M}$ on the right.} 
\label{Pathintegral}
\end{figure}

If we choose a different gauge
\begin{equation}
F_2(t):=\bar{\tau}(t)-t=0, \label{F2gauge}
\end{equation} 
in \eqref{pathint3}, we obtain 
\begin{align}
&\Delta_F(X_f; X_i|h_f ; h_i) \nonumber \\
&\quad
=Z_3 \int_{h_i X_i}^{h_f, X_f} 
\mathcal{D} h  \mathcal{D} X(\bar{\tau})
\mathcal{D} \alpha \mathcal{D}c \mathcal{D}b
\nonumber \\
&\qquad\quad
\exp \Biggl(- \int_{-\infty}^{\infty} dt \Bigl( +\alpha(t) (\bar{\tau}-t) +b(t)c(t)(1-\frac{d \bar{\tau}(t)}{dt})   \nonumber \\
&\qquad\qquad\qquad
+\int d\bar{\sigma} \sqrt{\bar{h}}G_{\mu\nu}(X(\bar{\tau}(t), t))  ( 
\frac{1}{2}\bar{h}^{00}\partial_{t} X^{\mu}(\bar{\tau}(t), t)\partial_{t} X^{\nu}(\bar{\tau}(t), t) +\bar{h}^{01}\partial_{t} X^{\mu}(\bar{\tau}(t), t)\partial_{\bar{\sigma}} X^{\nu}(\bar{\tau}(t), t) \nonumber \\
&\qquad\qquad\qquad\qquad\qquad\qquad
+\frac{1}{2}\bar{h}^{11}\partial_{\bar{\sigma}} X^{\mu}(\bar{\tau}(t), t)\partial_{\bar{\sigma}} X^{\nu}(\bar{\tau}(t), t)
)  \nonumber \\
&\qquad\qquad\qquad+\int d\bar{\sigma}\,i\,
B_{\mu\nu} (X(\bar{\tau}(t), t))
\partial_{t} X^{\mu}(\bar{\tau}(t), t)\partial_{\bar{\sigma}} X^{\nu}(\bar{\tau}(t), t) 
\nonumber \\
&\qquad\qquad\qquad
+\frac{1}{2}\int d\bar{\sigma}\sqrt{\bar{h}}\,
\alpha'\,R_{\bar{h}}
\Phi (X(\bar{\tau}(t), t))\Bigr) \Biggr) \nonumber \\
&\quad
=Z\int_{h_i, X_i}^{h_f, X_f} 
\mathcal{D} h  \mathcal{D} X(\bar{\tau})
\nonumber \\
&\qquad\quad
\exp \Biggl(- \int_{-\infty}^{\infty} d\bar{\tau} 
\int d\bar{\sigma} \sqrt{\bar{h}}G_{\mu\nu}(X(\bar{\sigma}, \bar{\tau}))  ( 
\frac{1}{2}\bar{h}^{00}\partial_{\bar{\tau}} X^{\mu}(\bar{\sigma}, \bar{\tau})\partial_{\bar{\tau}} X^{\nu}(\bar{\sigma}, \bar{\tau})
+\bar{h}^{01}\partial_{\bar{\tau}} X^{\mu}(\bar{\sigma}, \bar{\tau})\partial_{\bar{\sigma}} X^{\nu}(\bar{\sigma}, \bar{\tau}) 
 \nonumber \\
&\qquad\qquad\qquad\qquad\qquad\qquad\qquad
+\frac{1}{2}\bar{h}^{11}\partial_{\bar{\sigma}} X^{\mu}(\bar{\sigma}, \bar{\tau})\partial_{\bar{\sigma}} X^{\nu}(\bar{\sigma}, \bar{\tau}))
 \nonumber \\
&\qquad\qquad\qquad+\int d\bar{\sigma}\,i\,
B_{\mu\nu} (X(\bar{\sigma}, \bar{\tau}))
\partial_{\bar{\tau}} X^{\mu}(\bar{\sigma}, \bar{\tau})\partial_{\bar{\sigma}} X^{\nu}(\bar{\sigma}, \bar{\tau}) 
\nonumber \\
&\qquad\qquad\qquad
+\frac{1}{2}\int d\bar{\sigma}\sqrt{\bar{h}}\,
\alpha'\,R_{\bar{h}}
\Phi (X(\bar{\sigma}, \bar{\tau})) \Biggr). \label{prelastaction}
\end{align}
The path integral is defined over all possible two-dimensional Riemannian manifolds with fixed punctures in the manifold $\mathcal{M}$ defined by the metric $G_{\mu\nu}$, as in Fig. \ref{Pathintegral}.
The diffeomorphism times Weyl invariance of the action in \eqref{prelastaction} implies that the correlation function is given by 
\begin{equation}
\Delta_F(X_f; X_i|h_f ; h_i)
=
Z
\int_{h_i, X_i}^{h_f, X_f} 
\mathcal{D} h  \mathcal{D} X
e^{iS_{s}}, 
\label{FinalPropagator}
\end{equation}
where
\begin{align}
S_{s}
=
&\frac{1}{2}\int_{-\infty}^{\infty} d\tau \int d\sigma \sqrt{h(\sigma, \tau)}
\nonumber \\
&\hspace{15mm}
\times\biggl(\bigl( h^{mn} (\sigma, \tau) G_{\mu\nu}(X(\sigma, \tau))
+i \varepsilon^{mn}(\sigma, \tau)
B_{\mu\nu}(X(\sigma, \tau))\bigr) \partial_m X^{\mu}(\sigma, \tau) \partial_n X^{\nu}(\sigma, \tau) 
\nonumber \\
&\hspace{25mm}
+\alpha'\,R_{\bar{h}}\,\Phi(X(\sigma, \tau))\biggr). 
\end{align}
For regularization, we divide the correlation function by $Z$ and the volume of the diffeomorphism and the Weyl transformation $V_{diff \times Weyl} $, by renormalizing $\tilde{\phi}$.  \eqref{FinalPropagator} is the path integrals of   perturbative strings on an arbitrary background that possess the moduli in the string theory themselves \cite{Polchinski:1998rr}. Especially, in string geometry, the consistency of the perturbation theory around the background \eqref{Gh},  (\ref{Bdmu}),  (\ref{Bmunu}) and (\ref{Phi})   determines $d=26$ (the critical dimension).

\section{Conclusion and Discussion}
\setcounter{equation}{0}
In this paper, in the closed bosonic sector of string geometry theory, we fix the classical part of the scalar fluctuation of the metric around the string background configurations, which are parametrized by the string backgrounds, $G_{\mu\nu}(x)$, $B_{\mu\nu}(x)$, and $\Phi (x)$. We showed that the two-point correlation functions of the quantum parts of the scalar fluctuation are path integrals of the  perturbative strings  on the string backgrounds. In this derivation, we move from the second quantization formalism to the first one, where  the coordinates of the two fields in the correlation functions become the asymptotic fields that represent the initial state $X^{\mu}(\tau=-\infty, \sigma)$ and the final state $X^{\mu}(\tau=\infty, \sigma)$, respectively. All the paths on the string manifolds from $X^{\mu}(\tau=-\infty, \sigma)$ to  $X^{\mu}(\tau=\infty, \sigma)$ are summed up in the first quantization representation of the two-point correlation functions. Because the paths on the string manifolds are world-sheets with genera as shown in the section two in \cite{Sato:2017qhj}, they reproduce the  path integrals of the  perturbative strings up to any order, although the correlation functions are at tree level.

Next task is a supersymmetric generalization of our result.
It is known to be too difficult to describe the action of the perturbative strings on the R-R backgrounds in the NS-R formalism. 
Because string geometry theory is formulated in the NS-R formalism, we should derive the path integrals of the perturbative strings on the NS-NS backgrounds.

\section*{Acknowledgements}
We would like to thank  
H. Kawai,
K. Kikuchi,
T. Yoneya,
and especially 
A. Tsuchiya
for discussions.
The research of YS is supported by Samsung Science and Technology Foundation under Project Number SSTF-BA2002-05 and by the National Research Foundation of Korea (NRF) grant funded by the Korea government (MSIT) (No. 2018R1D1A1B07042934).

\appendix

\section{Green function on string geometry}
\setcounter{equation}{0}
In this appendix, we will show that  (\ref{GreenFunc}) is indeed a Green function on the flat string manifold. 
If $X^\mu(\bar{\sigma})
\not\equiv {X'}^\mu(\bar{\sigma})$,
we have 
\Eqr{
&&\frac{1}{\bar{e}'}
\frac{\pd}{\pd X_\nu(\bar{\sigma}')}\,{\cal N}\,
\left[\int d\bar{\sigma}
\frac{\bar{e}^2}{\sqrt{\bar{h}}}\left(X^\mu(\bar{\sigma})
-{X'}^\mu(\bar{\sigma})\right)^2
\right]^{\frac{2-D}{2}}\nn\\
&&~~~
=(2-D)\,{\cal N}\,
\left[\int d\bar{\sigma}
\frac{\bar{e}^2}{\sqrt{\bar{h}}}\left(X^\mu(\bar{\sigma})
-{X'}^\mu(\bar{\sigma})\right)^2
\right]^{-\frac{D}{2}}
\frac{\bar{e}'}{\sqrt{\bar{h}'}}\left(X^\nu(\bar{\sigma}')
-{X'}^\nu(\bar{\sigma}')\right),
}
and then, 
\Eqr{
&&\frac{1}{\bar{e}''}
\frac{\pd}{\pd X^\nu(\bar{\sigma}'')}
\frac{1}{\bar{e}'}
\frac{\pd}{\pd X_\nu(\bar{\sigma}')}\,{\cal N}\,
\left[\int d\bar{\sigma}
\frac{\bar{e}^2}{\sqrt{\bar{h}}}
\left(X^\mu(\bar{\sigma})
-{X'}^\mu(\bar{\sigma})\right)^2
\right]^{\frac{2-D}{2}}\nn\\
&&~~~
=d(2-D)\frac{1}{\sqrt{\bar{h}'}}\frac{\bar{e}'}{\bar{e}''}
\,{\cal N}\,
\left[\int d\bar{\sigma}\frac{\bar{e}^2}{\sqrt{\bar{h}}}
\left(X^\mu(\bar{\sigma})
-{X'}^\mu(\bar{\sigma})\right)^2
\right]^{-\frac{D}{2}}
\delta(\bar{\sigma}'-\bar{\sigma}'')\nn\\
&&~~~~~
-D(2-D)\frac{\bar{e}'}{\sqrt{\bar{h}'}}
\frac{\bar{e}''}{\sqrt{\bar{h}''}}
\,{\cal N}\,\left[\int d\bar{\sigma}
\frac{\bar{e}^2}{\sqrt{\bar{h}}}
\left(X^\mu(\bar{\sigma})
-{X'}^\mu(\bar{\sigma})\right)^2
\right]^{-\frac{D+2}{2}}\nn\\
&&~~~~~\times
\left(X^\nu(\bar{\sigma}')
-{X'}^\nu(\bar{\sigma}')\right)
\left(X_\nu(\bar{\sigma}'')
-{X'}_\nu(\bar{\sigma}'')\right)\,.
}
Thus, 
\Eqr{
&&\int d\bar{\sigma}'\sqrt{\bar{h}'}\frac{1}{\bar{e}'}
\frac{\pd}{\pd X^\nu(\bar{\sigma}')}
\frac{1}{\bar{e}'}
\frac{\pd}{\pd X_\nu(\bar{\sigma}')}\,{\cal N}\,
\left[\int d\bar{\sigma}
\frac{\bar{e}^2}{\sqrt{\bar{h}}}
\left(X^\mu(\bar{\sigma})
-{X'}^\mu(\bar{\sigma})\right)^2
\right]^{\frac{2-D}{2}}\nn\\
&&
=d\int d\bar{\sigma}' \delta(0)\,(2-D)\,{\cal N}\,
\left[\int d\bar{\sigma}
\frac{\bar{e}^2}{\sqrt{\bar{h}}}
\left(X^\mu(\bar{\sigma})
-{X'}^\mu(\bar{\sigma})\right)^2
\right]^{-\frac{D}{2}}\nn\\
&&~~~
-D(2-D)\,{\cal N}\,
\left[\int d\bar{\sigma}
\frac{\bar{e}^2}{\sqrt{\bar{h}}}
\left(X^\mu(\bar{\sigma})
-{X'}^\mu(\bar{\sigma})\right)^2
\right]^{-\frac{D+2}{2}}
\int d\bar{\sigma}'\,
\frac{\bar{e}'^2}{\sqrt{\bar{h}'}}
\left(X^\nu(\bar{\sigma}')
-{X'}^\nu(\bar{\sigma}')\right)^2\nn\\
&&
=0,
}
where we use $D=d\int d\bar{\sigma}' \delta(0)$.
Hence, we find
\Eqr{
&&\int d\bar{\sigma}' \sqrt{\bar{h}'} \frac{1}{\bar{e}'}
\frac{\pd}{\pd X^\nu(\bar{\sigma}')}
\frac{1}{\bar{e}'}
\frac{\pd}{\pd X_\nu(\bar{\sigma}')}
\,\mathcal{N}\,
\left[\int d\bar{\sigma}
\frac{\bar{e}^2}{\sqrt{\bar{h}}}
\left(X^\mu(\bar{\sigma})
-{X'}^\mu(\bar{\sigma})\right)^2
\right]^{\frac{2-D}{2}}\nn\\
&&~~~
=\delta(X-X'),
}
where $\mathcal{N}$ is a normalizing constant.

\end{document}